\documentclass[%
 reprint,
 amsmath,amssymb,
 aps, dvipsnames,
]{revtex4-2}

\usepackage{graphicx}
\usepackage{dcolumn}
\usepackage{bm}

\usepackage{siunitx}
\usepackage{braket}
\usepackage{natbib}
\usepackage{enumitem}
\usepackage{color}
\usepackage{comment}
\usepackage[hidelinks]{hyperref}

\newcommand{\mycomment}[1]{}

\newcommand{\ui}{\Vec{u}^{\,i}} 
\newcommand{\e}{\mathrm{e}}
\newcommand{\ii}[1]{_{\textrm{#1}}}
\newcommand{\alphainter}{\alpha\ii{inter}}
\newcommand{\alphamol}{\alpha\ii{mol}}

\begin{document}

\preprint{APS/123-QED}

\title{High-harmonic generation from two weakly coupled molecules: a simple tight-binding model} 

\author{Lina Bielke$^1$}
\author{Samuel Sch\"opa$^1$}
\author{Falk-Erik Wiechmann$^{1,2}$}
\author{Franziska Fennel$^{1,2}$}
\author{Dieter Bauer$^{1,2}$}

\affiliation{%
$^1$Institute of Physics, University of Rostock, 18051 Rostock, Germany.\\
$^2$Department Life, Light \& Matter, University of Rostock, 18051 Rostock, Germany.
}
\date{\today}

\begin{abstract}
The generation of high harmonics is a strongly nonlinear effect that allows to probe properties of the target and to study electron dynamics in matter. It has been investigated in many different kinds of targets, including molecular gases, liquids and solids. Recently, high-harmonic generation was studied in organic molecular crystals by Wiechmann et al. [Nat. Commun. 16, 9890 (2025)]. It was found that the laser-polarization-dependent harmonic yield is sensitive to the weak couplings between nearest- and next-nearest-neighbor molecules. In this paper, the impact of the laser polarization angle and the intermolecular interaction on the harmonic yield is examined in detail using a simple but insightful two-dimensional tight-binding system that models a molecular dimer, i.e. two weakly coupled molecules. We find that the intensities of lower harmonic orders tend to maximize for a laser polarization direction aligning with the molecular axes, whereas higher harmonic orders rather show the strongest yield for a polarization direction along the intermolecular axis. We further demonstrate that the harmonic order at which the maximum flips from the molecular to the intermolecular direction strongly depends on the intermolecular coupling strength. To gain a deeper insight into the origins of the findings, we include a detailed adiabatic analysis, showing that the flipping of the maximum yield towards the intermolecular direction is already contained qualitatively in the adiabatically following states.
\end{abstract}

\maketitle

\section{Introduction}
High-harmonic generation (HHG) is an important process in strong field physics, in which a target illuminated by an intense laser pulse emits light with frequencies equal to multiples of the incident laser frequency. This nonlinear effect enables to produce ultrashort laser pulses~\cite{Antoine1996, Paul2001, Goulielmakis2008} as well as to study target properties in an all-optical way~\cite{Ghimire2011, Luu2015, Vampa2015, Lakhotia2020}. 
HHG has been examined in many different target classes over the past decades, such as atomic \cite{McPherson1987, Ferray1988, Krause1992, Macklin1993} and molecular gases \cite{Lynga1996, Lein2002, Smirnova2009}, clusters \cite{Tisch1997}, semiconductor nanostructures \cite{Peschel2022}, liquids \cite{Luu2018, Neufeld2022}, inorganic crystals \cite{Ghimire2011, You2017}, two-dimensional materials \cite{Tancogne-Dejean2018, Jurss2021, Zhang2021}, topological insulators \cite{Schmid2021, Jurss2021} and metals \cite{Korobenko2021}.\\\indent
Recently, also high-order harmonic generation in organic molecular crystals (OMCs) has been demonstrated~\cite{Wiechmann2025}.
Organic molecular crystals consist of aligned, weakly coupled molecules, which are arranged on a crystal lattice. Thus, they combine molecular and crystal features, which makes them an exciting new target class that is also of interest for various applications, such as organic field-effect transistors (OFETs) \cite{Reese2007, Hasegawa2009, Lin1997} or photovoltaic cells \cite{Wilson2013, Congreve2013}. In \cite{Wiechmann2025}, it has been shown that the intermolecular coupling plays a crucial role for the high-harmonic generation process in these targets, although it is weak compared to the intramolecular coupling or the interatomic bondings in inorganic crystals \cite{Skobeltsyn1965, BookOrganicMolecularSolids2007}.
Wiechmann et al. observed harmonics up to the order $17$ in a pentacene single crystal, a prototypical OMC. It was found that the harmonic yield maximizes for laser polarization directions aligning with the (next-)nearest neighbor directions in the crystal. The experimental findings could be reproduced qualitatively with a simplified two-dimensional tight-binding model of the crystal, showing that the harmonic yield is extremely sensitive to the couplings between the molecules \cite{Wiechmann2025}.\\\indent
In this paper, we use a similar tight-binding model but consider only a dimer consisting of two weakly coupled molecules, as in the unit cell of polyacenes \cite{Bragg1921, Douglas2006, Campbell1962, Anthony2008}. We examine the dependence of the harmonic yield on the laser polarization direction and the intermolecular coupling strength in detail and demonstrate that the main effects observed in \cite{Wiechmann2025} can already be seen in this simple dimer model system. Due to its relative simplicity, the model allows to gain a deeper insight into the origins of the polarization and coupling dependence.
To this end, we also use an adiabatic analysis, which shows that most of the observed features can already be reproduced qualitatively by assuming that the electrons follow the laser field adiabatically.\\\indent
The work is structured as follows: In Section\,\ref{section:exact}, the model is first introduced theoretically (in \ref{subsection:exact_theory}), and then the exact results for the laser polarization dependence are presented and analyzed (in \ref{subsection:exact_polarization_dependence}). Subsequently, the dependence on the intermolecular coupling strength is examined in Subsection\,\ref{subsection:exact_coupling_dependence}. Section\,\ref{section:adiabatic} focuses on the adiabatic treatment. After explaining the theory (\ref{subsection:adiabatic_theory}), the harmonic yield and its laser polarization dependence are studied (\ref{subsection:adiabatic_polarization_dependence}), followed by an investigation of the impact of the intermolecular coupling strength in the approximated model (\ref{subsection:adiabatic_coupling_dependence}).
A brief summary of the results is given in Section\,\ref{section:conclusion} along with an outlook on future research.

\section{Exact HHG calculations} \label{section:exact}
\subsection{Introduction of the tight-binding model system} \label{subsection:exact_theory}
\begin{figure}[h!]
    \centering
    \includegraphics[width=89mm]{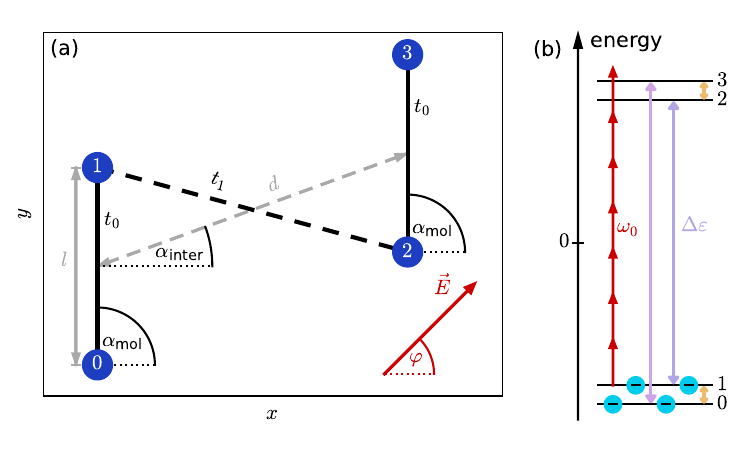}
    \caption{(a)~Sketch of the model system. The blue circles depict the tight-binding sites. The sites $0$ and $1$ ($2$ and $3$) form a molecule with the intramolecular coupling $t_0$. The molecules are connected by the intermolecular coupling $t_1$. The distances $l$ and $d$ and the angles $\alpha_{\textrm{mol}}$ and $\alpha_{\textrm{inter}}$ specify the geometry of the target. The electric laser field with the polarization direction $\varphi$ is indicated by the red arrow. (b)~Schematic representation of the energy levels of the unperturbed target system. Note that for the default settings, the splitting between the two lower (upper) levels is even smaller than depicted. In the ground state, the two lowest energy levels are doubly populated. The red arrows illustrate the energy of the fundamental laser photons. The dipole-allowed transitions are indicated by the purple and orange arrows.} 
    \label{fig:model}
\end{figure} 
In order to gain a better understanding of HHG in systems of weakly coupled molecules, we set up a simple tight-binding model, illustrated in Fig.\,\ref{fig:model}(a). The model simulates two parallel molecules. Each molecule consists of two sites, which are coupled via an intramolecular hopping amplitude $t_0$ (solid black lines). Additionally, the two closest sites of the different molecules are connected with the intermolecular hopping amplitude $t_1$ (dashed black line). For convenience, we set $t_0=-1$, defining the energy unit of the system. As we are interested in weakly coupled molecules, the default value of the intermolecular coupling is chosen significantly weaker, as 2\,\% of the intramolecular coupling, i.e. $t_1=-0.02$. 
Note that the minus sign in front of the coupling strengths is used by convention to obtain a ground state wave function without nodes, but it does not matter for the physical observables, since the energy levels of the system are arranged symmetrically around zero.\\\indent
The intermolecular distance was set to $d = 9.2$\,au, corresponding to a typical separation between neighboring molecules in OMCs. The molecular length was chosen as $l = 5.5$\,au, so that the ratio of inter- and intramolecular distance is comparable to that of real crystals.
We further define the unit system by using $\hbar = |e| = 1$ (as in atomic units) throughout the paper.\\\indent
For the angle of the molecular axes relative to the positive $x$-axis, $\alpha_{\textrm{mol}}=90$\,\si{\degree} is used. The angle of the intermolecular axis $\alpha_{\textrm{inter}}$, defined as the angle between the positive $x$-axis and the connecting line between the centers of the two molecules (see Fig.\,\ref{fig:model}(a)), is set to $20$\,\si{\degree}, such that the molecular and intermolecular direction are well separated but not orthogonal to each other.\\\indent
The eigenenergies $\varepsilon_i$ and eigenstates $\Vec{u}_i$ of the unperturbed system are calculated by solving the time-independent Schrödinger equation
\begin{align}
    \textbf{H}_0 \Vec{u}_i = \varepsilon_i \Vec{u}_i, \qquad i=0,1,2,3,
\end{align}
with the tight-binding Hamiltonian
\begin{align}
    \textbf{H}_0 = \begin{pmatrix}
        0 & t_0 & 0 & 0 \\
        t_0 & 0 & t_1 & 0 \\
        0 & t_1 & 0 & t_0 \\
        0 & 0 & t_0 & 0
    \end{pmatrix}.
\end{align}
We consider four electrons, corresponding to half filling. In the ground state, they populate the two lowest-energy, spin-degenerate eigenstates as depicted in the exemplary level scheme in Fig.\,\ref{fig:model}(b).\\\indent
In order to generate high harmonics, the model target is excited by a linearly polarized laser pulse in dipole approximation with the vector potential
\begin{align}
     \Vec{A}(t) = \frac{E_0}{\omega_0} \begin{pmatrix} \cos(\varphi) \\ \sin(\varphi) \end{pmatrix} \sin^2\left(\frac{\omega_0 t}{2 n_{\textrm{cyc}}} \right) \cos(\omega_0 t), \label{eq:A(t)}
\end{align}
for $0 \leq t \leq \frac{2 \pi n_{\textrm{cyc}}}{\omega_0}$ and $n_{\textrm{cyc}}=15$ laser cycles. The electric field follows as
\begin{align}
    \Vec{E}(t) = - \partial_t \Vec{A}(t).
\end{align}
The amplitude of the electric field is set to $E_0=0.15$ in the unit system of the model, yielding spectra with clear harmonics.
By default, the fundamental frequency $\omega_0$ is chosen such that $6.3$ laser photons correspond to the energy of the HOMO-LUMO gap $\Delta \varepsilon = \varepsilon_2 - \varepsilon_1$, i.e. $\omega_0 = \Delta \varepsilon/6.3$, similar to the experimental conditions in \cite{Wiechmann2025}. The default values are used throughout the paper unless stated otherwise.
The laser polarization direction, described by the angle $\varphi$, as illustrated in Fig.\,\ref{fig:model}(a), is varied throughout this work to analyze its impact on the high-harmonic generation.\\\indent
We couple the system to the laser pulse in length gauge. The resulting time-dependent Schrödinger equation reads
\begin{align}
    \mathrm{i} \partial_t \ui(t) = (\textbf{H}_0 + \textbf{H}_{\textrm{laser}}(t)) \ui(t) \label{eq:TDSE}
\end{align}
with the laser Hamiltonian
\begin{align}
    \textbf{H}_{\textrm{laser}}(t) = \Vec{E}(t) \cdot \Vec{\textbf{r}} = E_x(t)\textbf{X} + E_y(t)\textbf{Y}, \label{eq:H_laser}
\end{align}
where $\textbf{X}$ ($\textbf{Y}$) is a matrix with the $x$-($y$-)coordinates of the sites on the diagonal, i.e. $\textbf{X}=\textrm{diag}(x_0,x_1,x_2,x_3)$ and $\textbf{Y}=\textrm{diag}(y_0,y_1,y_2,y_3)$.\\\indent
The time evolution of the electronic wave functions $\ui(t)$ is calculated by solving \eqref{eq:TDSE} numerically using the matrix exponential and operator splitting:
\begin{align}
    \ui(t+dt) = \e^{-\mathrm{i} \textbf{H}_0 dt /2} \e^{-\mathrm{i} \textbf{H}_{\textrm{laser}}(t+dt/2) dt}  \e^{-\mathrm{i} \textbf{H}_0 dt /2} \ui(t),
\end{align}
with the ground state as initial condition and the time step $dt=0.1$.\\\indent
The position expectation value of an electron initially populating the eigenstate $\Vec{u}_i$ follows as
\begin{align}
    \langle \Vec{r}^{\,i} \rangle (t) = \begin{pmatrix}
        \Vec{u}^{\,i \dagger}(t) \textbf{X} \ui(t)\\
        \Vec{u}^{\,i \dagger}(t) \textbf{Y} \ui(t)
    \end{pmatrix}.
\end{align}
The total acceleration of all electrons
\begin{align}
    \langle \Vec{a} \rangle (t) =  \sum\limits\ii{el} \frac{d^2}{dt^2} \langle \Vec{r}^{\,i(\textrm{el})} \rangle (t)
\end{align}
is fast Fourier transformed, using a hann window, to calculate the harmonic yield 
\begin{align}
    I(\omega) = \left| \textrm{FFT}\left\{\langle \Vec{a} \rangle (t) \cdot \textrm{hann}(t)\right\} \right|^2.
    \label{eq:I(w)}
\end{align}
In order to obtain the intensity of a harmonic order $n$, we integrate the yield over a small frequency interval centered around the harmonic order according to
\begin{align}
    I_n = \int\limits_{(n-0.25)\omega_0}^{(n+0.25)\omega_0} I(\omega) \mathrm{d}\omega. \label{eq:I_n}
\end{align}
The integration limits are chosen such that we get representative values for the harmonics, avoiding the integration over artifacts and resonances.

\subsection{Dependence of the high-harmonic yield on the laser polarization direction} \label{subsection:exact_polarization_dependence}
\begin{figure}[h!]
    \centering
    \includegraphics[width=89mm]{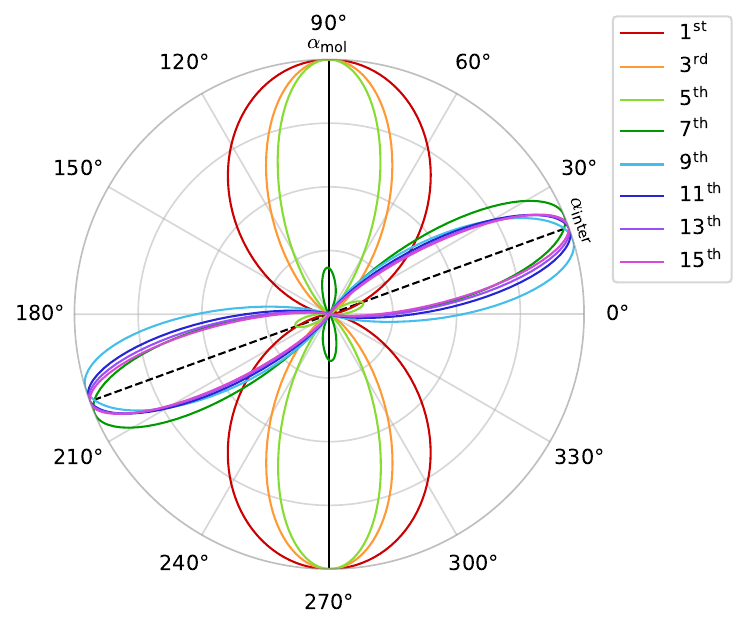}
    \caption{Normalized harmonic intensities as a function of the laser polarization direction $\varphi$. The black solid (dashed) line indicates the angle of the molecular (intermolecular) axis.}
    \label{fig:polar_basic}
\end{figure}

\begin{figure}[h!]
    \centering
    \includegraphics[width=89mm]{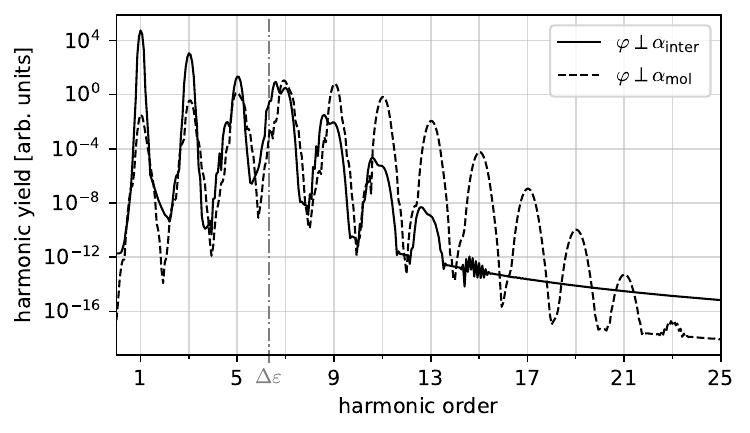}
    \caption{High-harmonic spectra for laser polarization directions perpendicular to the intermolecular (solid line) and molecular (dashed line) direction. The vertical dash-dotted gray line marks the frequency corresponding to the HOMO-LUMO gap of the unperturbed system.}
    \label{fig:spectra_basic}
\end{figure}

To analyze the high-harmonic generation in the described model system and to better understand the impact of the laser polarization direction on the harmonic yield, we calculate the intensities $I_n$ of the odd harmonics up to order $n=15$ with \eqref{eq:I_n} as a function of the laser polarization angle $\varphi$.
The corresponding results are depicted in Fig.\,\ref{fig:polar_basic}, where the curves are normalized by the maximum value for each harmonic individually, i.e. $I_n(\varphi)/[I_n(\varphi)]_{\textrm{max}}$.\\\indent
We observe that the lower harmonics (order $1$ to $5$) maximize when the laser is polarized parallel to the molecular axes, as marked by the solid black line.
The higher harmonic orders ($n \geq 7$), in contrast, show the highest yield for polarization angles close to $\alphainter$. The precise angles of the maxima vary slightly around $\alphainter$, as they depend on the details of the probability density distribution and the transition matrix elements. However, the deviations are small (less than $4$\,\si{\degree}).
It thus seems that there are two important contributions to the harmonic yield: a molecular contribution that leads to strong low-order harmonics if the electric laser field has a large component in the molecular direction, and an intermolecular contribution for laser polarizations where the electrons are driven in the intermolecular direction, which dominates for the higher orders.
For the $5^{\textrm{th}}$ and $7^{\textrm{th}}$ harmonic, a second, smaller lobe appears in the respective other characteristic direction, indicating that the molecular and intermolecular contribution are on a similar order of magnitude for these harmonics.\\\indent
The results match the findings in Wiechmann et al. \cite{Wiechmann2025} for a periodic tight-binding model of an organic molecular crystal, where the same effect of the lobes flipping from the molecular to the intermolecular direction at some harmonic order was observed.
Also the narrowing of the lobes with increasing harmonic order, which was seen in the measurements and theoretical calculations of \cite{Wiechmann2025}, is reproduced by our dimer model.
This agreement with the results for HHG in organic molecular crystals supports the hypothesis that the response of the crystals is mainly governed by electron population transfers within the molecules and between neighboring molecules. Furthermore, it shows that our simple model captures the essential features to describe HHG in systems of weakly coupled molecules. 
Compared to a crystal model, the system with only two molecules is much easier to analyze, allowing us to understand the observed behavior in more detail.\\\indent 
To this end, it is first of all insightful to take a look at harmonic spectra for a laser polarization perpendicular to the intermolecular axis described by $\alphainter$ (i.e. $\varphi = 110$\,\si{\degree}) and for a laser polarization perpendicular to the molecular axes (i.e. $\varphi = 0$\,\si{\degree}). These spectra are plotted in Fig.\,\ref{fig:spectra_basic} and denoted by $\varphi \perp \alphainter$ and $\varphi \perp \alphamol$, respectively.
By choosing laser fields with a polarization direction perpendicular to the characteristic axes of the target, we effectively separate the molecular and intermolecular response to the laser pulse.\\\indent
Both spectra show clear odd harmonics, whereas no even harmonics occur as the system is inversion symmetrical.
However, the characteristics of the two spectra are very different.
For $\varphi \perp \alphainter$, the yield is high for the lowest orders, but decreases strongly with increasing harmonic order. In contrast, the spectrum for $\varphi \perp \alphamol$ shows the highest yield for harmonics close to the frequency corresponding to the HOMO-LUMO gap, which is indicated by the vertical dash-dotted line. With increasing spectral distance to $\Delta \varepsilon$, the harmonic yield decreases. This suggests that transitions play an important role for the harmonic generation process of the intermolecular contribution.\\\indent
When looking at the dipole-allowed transitions in our system, we find that the transitions between $\varepsilon_0 \leftrightarrow \varepsilon_1$ and $\varepsilon_2 \leftrightarrow \varepsilon_3$ (orange arrows in Fig.\,\ref{fig:model}(b)) have a large dipole moment, which points in the intermolecular direction, i.e. $20$\,\si{\degree}. The higher-energy transitions $\varepsilon_1 \leftrightarrow \varepsilon_2$ and $\varepsilon_0 \leftrightarrow \varepsilon_3$ (purple arrows in Fig.\,\ref{fig:model}(b)), in contrast, have a dipole moment pointing in $-90.9$\,\si{\degree} and $-89.1$\,\si{\degree} direction, respectively, so almost along the molecular axes.
One could naively expect that these higher-energy transitions therefore do not play an important role for the high-harmonic generation with $\varphi \perp \alphamol$. However, they are essential, as without them, no electron population would be in the higher states to drive $\varepsilon_2 \leftrightarrow \varepsilon_3$, and contributions from $\varepsilon_0\leftrightarrow \varepsilon_1$ would be prevented by the Pauli exclusion principle.
The spectrum shows that these multi-photon processes with energies around $\Delta \varepsilon$ are indeed crucial for the intermolecular contribution.\\\indent
For other polarization directions than the depicted, we see a combination of the two contributions, weighted according to the angular distance of $\varphi$ to $\alphamol$ and $\alphainter$.
The distinct characteristics of the molecular and intermolecular HHG contribution explain the observed flipping of the yield-maximizing polarization angle from $\varphi = \alphamol$ for lower harmonic orders to $\varphi \approx \alphainter$ for the higher orders.

\subsection{Influence of the intermolecular coupling strength} \label{subsection:exact_coupling_dependence} 
\begin{figure*}[ht]
    \centering
    \includegraphics[width=183mm]{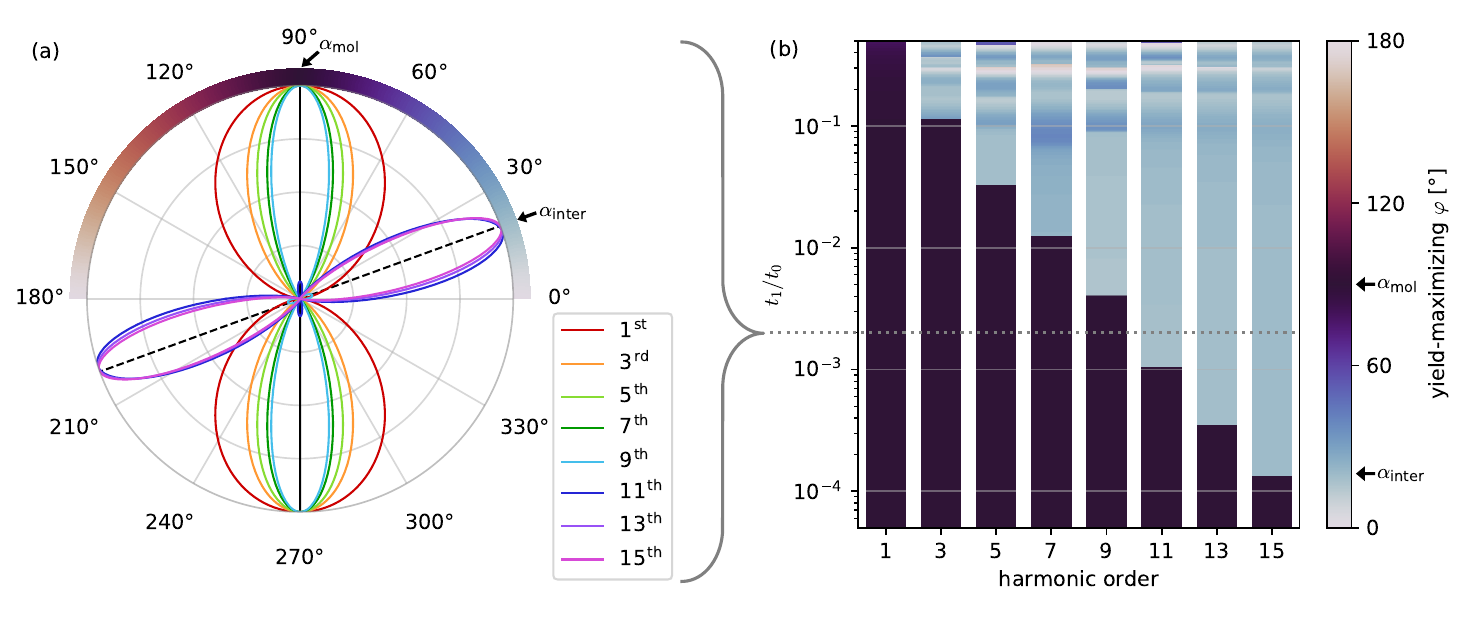}
    \caption{(a)~Normalized harmonic intensities as a function of the laser polarization angle for the reduced intermolecular coupling of $t_1=0.002\,t_0$.
    (b)~Yield-maximizing laser polarization angle for the different harmonic orders color-coded as a function of the intermolecular coupling strength $t_1$ (given in units of the intramolecular coupling strength $t_0$). The light blue corresponds to polarization angles close to $\alphainter$, whereas the dark shades indicate polarization directions close to the orientation of the molecular axes, as also illustrated in panel (a).}
    \label{fig:polar_and_peakphis_exact}
\end{figure*}

\begin{figure}[h!]
    \centering
    \includegraphics[width=89mm]{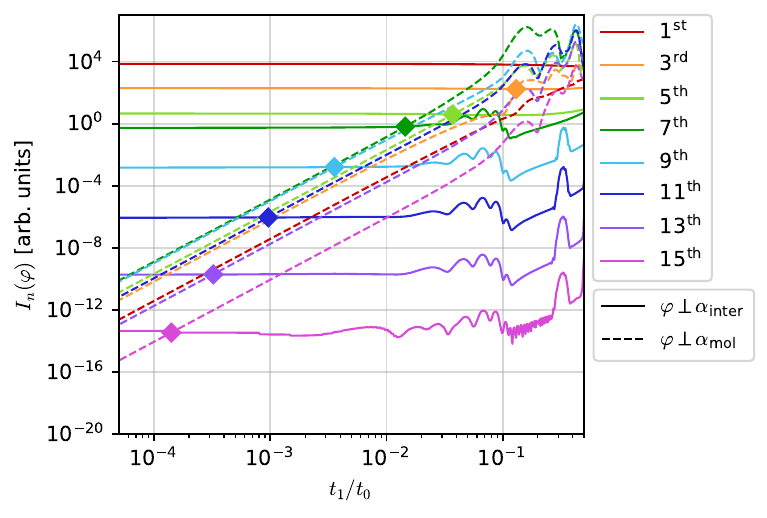}
    \caption{Harmonic intensities as a function of the intermolecular coupling strength $t_1$ for a laser polarization perpendicular to the intermolecular axis (solid lines) and perpendicular to the molecular axes (dashed lines). The diamonds indicate where the yield for $\varphi \perp \alphamol$ starts to exceed the yield for $\varphi \perp \alphainter$.}
    \label{fig:t1-dep_I_n}
\end{figure}

Now we want to elaborate on the impact of the intermolecular coupling strength $t_1$ on the high-harmonic generation in our model system of weakly coupled molecules. 
Figure\,\ref{fig:polar_and_peakphis_exact}(a) shows the polarization-dependent harmonic yield, similar to Fig.\,\ref{fig:polar_basic}, but for a ten times weaker intermolecular coupling strength of $t_1=0.002\,t_0$.
We find that the principle features are retained for the reduced coupling: again lower harmonic orders show a lobe around $\varphi = \alphamol$, whereas the highest depicted orders maximize for a polarization parallel to the intermolecular direction. However, due to the lower coupling, the flipping of the maximum to the intermolecular angle only occurs for higher harmonic orders, more specifically, from the $11^{\textrm{th}}$ harmonic onwards, rather than from the $7^{\textrm{th}}$ (as observed for $t_1=0.02\,t_0$). This demonstrates that the intermolecular coupling strength strongly influences the polarization dependence of the harmonic yield.\\\indent
In order to analyze the impact in more detail, the intermolecular coupling strength is varied over several orders of magnitude and the polarization-dependent harmonic intensities are calculated as a function of $t_1$.
In Fig.\,\ref{fig:polar_and_peakphis_exact}(b), we display the yield-maximizing laser polarization direction color-coded for the different harmonic orders and coupling strengths. The dark color means that the harmonic intensity is highest for a laser polarization along the molecular axes, the light blue indicates that the harmonic yield is maximized for polarization angles around the intermolecular direction.
The plot reveals that for very weak intermolecular couplings, around four orders of magnitude smaller than the intramolecular coupling, all depicted harmonics exhibit the highest intensity for a laser pulse that is polarized parallel to the molecular axes. When increasing $|t_1|$, first the $15^{\textrm{th}}$, so the highest of the considered harmonic orders, shows a flipping of the yield-maximizing polarization angle to the intermolecular direction $\varphi \approx \alphainter$. Notably, this already happens for a quite weak intermolecular coupling of $t_1 \approx 0.02\%\,t_0$.
By further increasing the strength of the intermolecular coupling, one can observe the flipping also for the other harmonics from the $13^{\textrm{th}}$ to the $3^{\textrm{rd}}$, one after another.\\\indent
In the regime of $|t_1| \gtrsim 0.1\,|t_0|$, the yield-maximizing polarization angles show larger fluctuations around $\alphainter$.
At such strong couplings, the model loses physical interpretability, as too much electron population is excited in our limited four-level scheme.
Nevertheless, the figure reveals a clear dependence of the polarization-dependent harmonic intensities on the intermolecular coupling strength.
This effect may pave the way towards an all-optical method to probe the coupling strength between molecules, as it was proposed in \cite{Wiechmann2025} for OMCs, where a qualitatively similar impact of the intermolecular coupling strength on the harmonic emission for a crystal was found.\\\indent
But how can the observed behavior be explained?
It is intuitive that a laser pulse polarized parallel to the intermolecular axis can more effectively generate high harmonics when the intermolecular coupling increases.
It is, however, less obvious why higher harmonic orders show the described flipping already for weaker intermolecular couplings than lower orders. Also, the flipping as a function of the harmonic order and $t_1$ appears almost linear in the logarithmic plot.
To understand this regularity, it is helpful to again consider the cases where the laser is polarized perpendicular to the characteristic axes of the target. The harmonic intensities for these two polarization angles are plotted as a function of the intermolecular coupling strength in Fig.\,\ref{fig:t1-dep_I_n}.\\\indent
For the polarization perpendicular to the intermolecular direction (solid lines), we observe two main features: First, the harmonic intensities are almost independent of $t_1$, as long as it is significantly weaker than the intramolecular coupling. And second, the yields for the different harmonic orders are about equally spaced in the logarithmic plot and sorted in sequence, matching the observation from the spectrum in Fig.\,\ref{fig:spectra_basic} that the yield drops roughly exponentially with the harmonic order for this polarization direction. Larger deviations from the equal spacing occur close to the resonance, especially for the $7^{\textrm{th}}$ harmonic.\\\indent
In contrast, the harmonic intensities for $\varphi \perp \alphamol$ (dashed lines) exhibit a clear increase with the intermolecular coupling strength up to $t_1 \approx 0.1\,t_0$. We find that all harmonics show the same slope according to a scaling of the intensities with $|t_1|^4$.
This relation can be traced back to the scaling of other variables with $t_1$: For the laser polarization direction perpendicular to the molecular axes, the accelerations of the individual electrons show leading terms proportional to $|t_1|$ (for $t_1/t_0 \ll 1$). However, when summing over all electronic contributions, the terms linear in $t_1$ destructively interfere and consequently, the first term contributing to the total acceleration $\langle \Vec{a} \rangle (t)$ is quadratic in the intermolecular coupling strength. One can indeed easily proof that the contributions linear in $t_1$ cancel out for the transition dipole moments pointing in the intermolecular direction (i.e. for $\varepsilon_0 \leftrightarrow \varepsilon_1$ and $\varepsilon_2 \leftrightarrow \varepsilon_3$) by performing a Taylor expansion of them around $t_1/t_0=0$. Since the harmonic intensities are related to the square of the total electron acceleration, they show a leading term proportional to $|t_1|^4$.\\\indent
As seen in Fig.\,\ref{fig:spectra_basic}, the yield for $\varphi \perp \alphamol$ is highest for the harmonics close to the HOMO-LUMO gap of $6.3\,\omega_0$. However, the intensities of the different harmonic orders are much closer to each other for this polarization direction than for $\varphi \perp \alphainter$, resulting in closer-lying curves in Fig.\,\ref{fig:t1-dep_I_n}.
These different dependencies explain the behavior observed in Fig.\,\ref{fig:polar_and_peakphis_exact}(b): For very weak intermolecular couplings the molecular contribution dominates for all harmonics.
With increasing $|t_1|$, the intermolecular contribution rises and starts to dominate the highest harmonic first, here the $15^{\textrm{th}}$, and then the lower orders one after another, as marked in Fig.\,\ref{fig:t1-dep_I_n} by diamonds in the respective colors. 
Due to the approximately equal spacing of the solid lines and the similar behavior of the different dashed lines, the crossing points appear at roughly regularly spaced $t_1$-values in the logarithmic plot.

\section{Adiabatic analysis} \label{section:adiabatic}
\subsection{Theory} \label{subsection:adiabatic_theory}
In order to gain an even deeper understanding of the processes relevant to the high-harmonic generation in the system of weakly coupled molecules, we perform an adiabatic analysis.\\\indent 
The adiabatic theorem states that if the changes in the laser field are slow compared to the dynamics in the system, the electrons can be treated as adiabatically following the laser pulse \cite{Born1928}. To this end, instead of using the time-dependent Schrödinger equation, the time-\textit{in}dependent Schrödinger equation is solved at each time step with the instantaneous Hamiltonian,
\begin{align}
   (\textbf{H}_0 + \textbf{H}\ii{laser}(t)) \Vec{\mu}_n(t) = \epsilon_n(t) \Vec{\mu}_n(t), \quad n=0,1,2,3,
\end{align}
yielding the instantaneous eigenstates $\Vec{\mu}_n(t)$ and eigenenergies $\epsilon_n(t)$.\\\indent
Instead of directly using this approximation, we expand the exact states in the basis of these adiabatic eigenstates according to
\begin{align}
    \ui(t) = \sum_{n=0}^3 c_n^i(t) \Vec{\mu}_n(t),
\end{align}
with the coefficients being calculated via 
\begin{align}
    c_m^i (t) = \Vec{\mu}_m^{\,\dagger} (t) \ui (t).
\end{align}
Inserting the expansion in the formula for the $x$-component of the position expectation value summed over all electrons gives
\begin{align}
    \langle x \rangle (t) 
    &= \sum\ii{el} \Vec{u}^{\,i(\textrm{el})\dagger}(t) \textbf{X} \Vec{u}^{\,i(\textrm{el})}(t)\\
    &= \sum_{m,n=0}^3 \underbrace{\sum\ii{el} c_m^{i(\textrm{el})*}(t) c_n^{i(\textrm{el})}(t)}_{C_{mn}(t)} \underbrace{\Vec{\mu}_m^{\,\dagger}(t) \textbf{X} \Vec{\mu}_n(t)}_{\xi_{mn}(t)}\\
    &= \sum_{m,n=0}^3 C_{mn}(t) \xi_{mn}(t). \label{eq:Cxi}
\end{align}
The calculation for the $y$-component is completely analogous. We call $\eta_{mn}(t) = \Vec{\mu}_m^{\,\dagger}(t) \textbf{Y} \Vec{\mu}_n(t)$ and $\Vec{r}_{mn}(t) = (\xi_{mn}(t), \eta_{mn}(t))^\top$.\\\indent
Now we want to consider two different approximations to examine the origin of the observed features in more detail:
\begin{enumerate}
    \item The electrons remain in the instantaneous eigenstates, i.e. they follow the laser field adiabatically.
    \item Only terms describing transitions between initially populated and initially unpopulated instantaneous eigenstates are considered.
\end{enumerate}
In condensed matter HHG, one typically distinguishes between intraband contributions to the harmonic yield (due to electronic movements within the energy bands) and interband contributions (due to transitions between energy bands). In analogy to this, we will refer to the first approximation, where the electrons stay within the adiabatic eigenstates, as "adia-intra" and to the second, where transitions between the energy levels are considered, as "adia-inter".\\\indent
The position expectation values for these approximations read
\begin{align}
    \langle \Vec{r} \rangle\ii{adia-intra} (t) &= 2 \Vec{r}_{00}(t) + 2 \Vec{r}_{11}(t) \quad \textrm{and}\label{eq:adia_1}\\
    \langle \Vec{r} \rangle\ii{adia-inter} (t) &= \sum_{m=0}^1 \sum_{n=2}^3 C_{mn}(t) \Vec{r}_{mn}(t) + \textrm{h.c.}.
\end{align}
We use these formulas instead of the exact position expectation value in the calculation of the harmonic yield described in Subsection\,\ref{subsection:exact_theory} to receive the approximated HHG results.

\subsection{Approximated high-harmonic yield and its dependence on the laser polarization direction} \label{subsection:adiabatic_polarization_dependence}
\begin{figure}[h!]
    \centering
    \includegraphics[width=89mm]{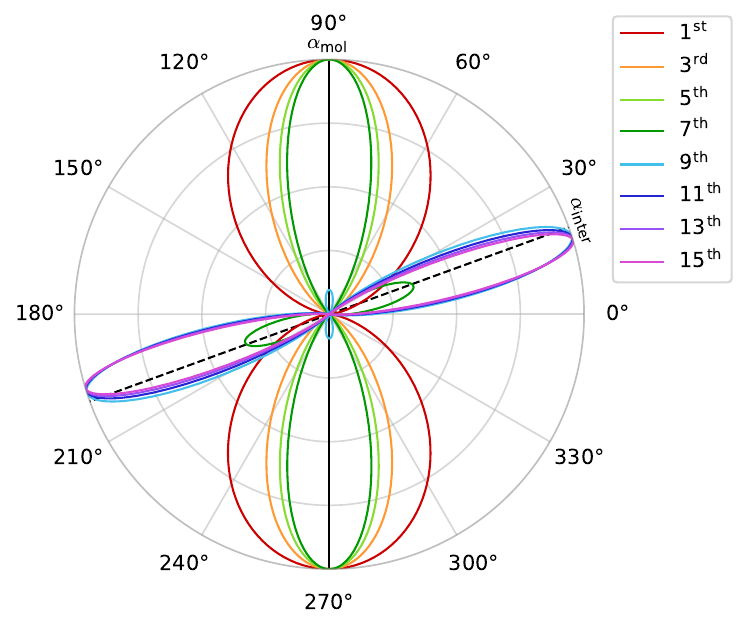}
    \caption{Normalized harmonic intensities as a function of the laser polarization angle as in Fig.\,\ref{fig:polar_basic}, but using the adiabatic approximation "adia-intra", i.e. the assumption that the electrons follow the laser pulse adiabatically.}
    \label{fig:ad_polar}
\end{figure}

To understand what the different approximations can account for, we first compare an exact spectrum with the corresponding harmonic spectra calculated with the two approximations described above. For the laser polarization angle, we choose $\varphi = 55$\,\si{\degree} to see both molecular and intermolecular features in the spectrum. Furthermore, we reduce the laser frequency $\omega_0$ to $\Delta \varepsilon/12.6$, i.e. to half the default frequency, for a better visibility of the features. 
\begin{figure}[h!]
    \centering
    \includegraphics[width=89mm]{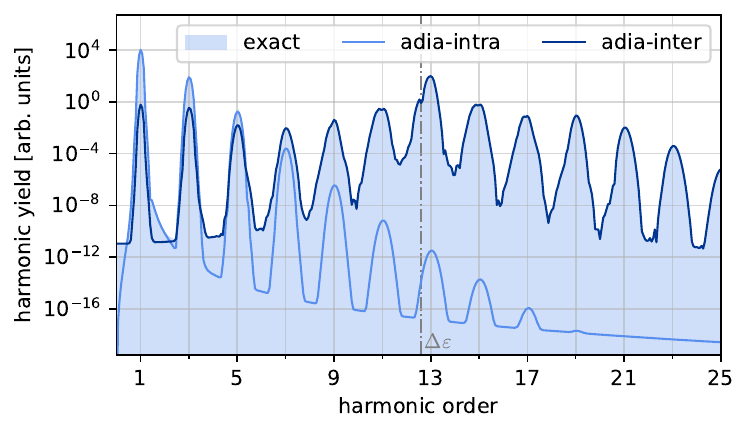}
    \caption{Exact harmonic spectrum along with the results for the two adiabatic approximations (specified in Subsection\,\ref{subsection:adiabatic_theory}) for $\varphi = 55$\,\si{\degree} and $\omega_0 = \Delta \varepsilon/12.6$.}
    \label{fig:ad_spectra}
\end{figure} 
As can be seen in Fig.\,\ref{fig:ad_spectra}, the lower laser frequency leads to a shift of the bump around the HOMO-LUMO gap (indicated by the vertical dash-dotted line) to higher harmonic orders, as expected. The exact spectrum now clearly shows a rapid decline of the emitted intensity with the harmonic order for the lower harmonics (as observed for the molecular contribution) combined with the high yield for harmonics close to $\Delta \varepsilon$, known from the intermolecular contribution. 
The approximation "adia-intra", i.e. the assumption that the electrons follow the instantaneous eigenstates adiabatically, leads to a spectrum with a rapidly decreasing harmonic yield, showing a good agreement with the exact result for the lower orders, whereas the bump around the HOMO-LUMO gap does not appear in the approximated spectrum.\\\indent
In contrast, the spectrum for the approximation "adia-inter", which only considers transitions from an instantaneous energy level below to a level above the HOMO-LUMO gap and vice versa, underestimates the yield for the $1^{\textrm{st}}$ to the $5^{\textrm{th}}$ harmonic, while the higher-order yield is reproduced correctly. This shows that the bump around the frequency corresponding to the HOMO-LUMO gap can be explained by transitions between the adiabatic eigenstates.\\\indent
For the default laser frequency, the principle observations are similar, but as the HOMO-LUMO gap corresponds to a lower harmonic order, the deviations of the "adia-intra" approximation from the exact results become apparent already at lower harmonic frequencies. This is also expected, as the approximation of adiabatically following electrons becomes worse for higher laser frequencies.\\\indent
However, we want to examine in how far the adiabatic approximation "adia-intra" is able to reproduce the observed polarization dependence of the harmonic intensities, as well as the impact of the intermolecular coupling strength, despite the shortcomings for the yield of the higher harmonics.
To this end, we repeat the calculations performed for Section\,\ref{section:exact} with the default laser frequency, replacing the exact position expectation value with that defined in~\eqref{eq:adia_1}.\\\indent
The dependence of the approximated harmonic intensities on the laser polarization direction is depicted in Fig.\,\ref{fig:ad_polar} analogous to Fig.\,\ref{fig:polar_basic}.
One can observe lobes around $\varphi = \alphamol$ for lower harmonics and maxima close to $\alphainter$ for higher harmonics, similar to the exact results. Also the decrease in the width of the lobes with increasing harmonic order is clearly visible. However, the $7^{\textrm{th}}$ harmonic exhibits its maximum in the molecular direction and has a smaller lobe in the intermolecular direction, as opposed to the exact case. 
Thus, quantitative deviations from the exact results are apparent, but the adiabatic approximation is able to reproduce the harmonic yield as a function of the incoming pulse's polarization surprisingly well.

\subsection{Influence of the intermolecular coupling strength in the adiabatic approximation} \label{subsection:adiabatic_coupling_dependence}
\begin{figure}[h!]
    \centering
    \includegraphics[width=89mm]{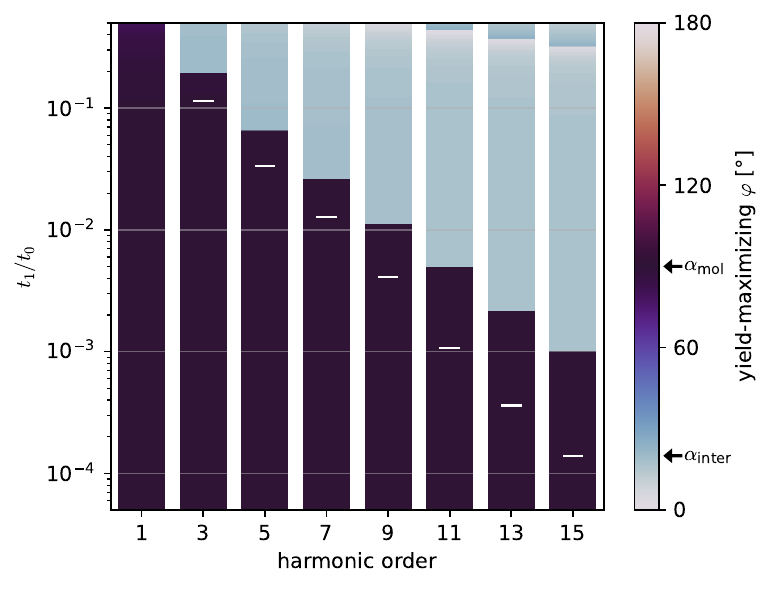}
    \caption{Laser polarization angle that maximizes the different harmonic intensities color-coded as a function of the intermolecular coupling strength, as in Fig.\,\ref{fig:polar_and_peakphis_exact}(b), but for adiabatically following electrons (approximation "adia-intra"). The white dashes mark the $t_1$-values for which the yield-maximizing polarization angle flips from the molecular to the intermolecular direction in the exact calculations.}
    \label{fig:ad_peakphis}
\end{figure}

\begin{figure}[h!]
    \centering
    \includegraphics[width=89mm]{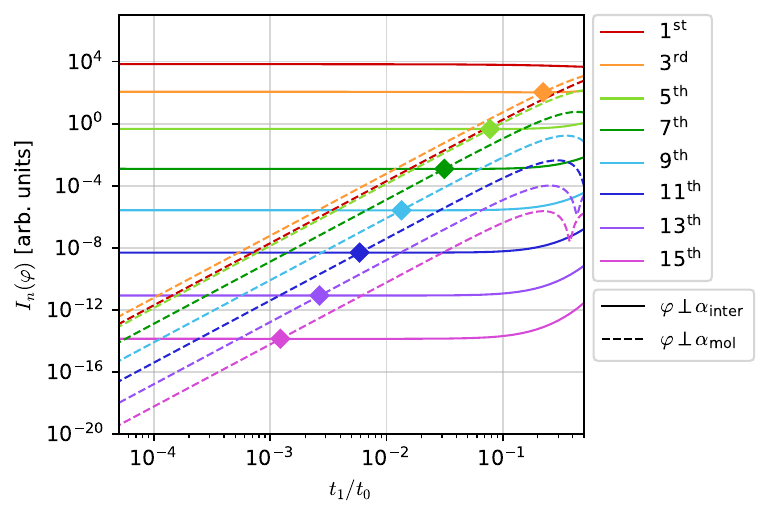}
    \caption{Intensities of the different harmonic orders for the laser polarization angles $\varphi \perp \alphainter$ and $\varphi \perp \alphamol$ as a function of the intermolecular coupling strength, as in Fig.\,\ref{fig:t1-dep_I_n}, but for the adiabatic approximation "adia-intra" (i.e. electrons staying in the instantaneous eigenstates). The diamonds again highlight the crossing points of the curves in the respective color.}
    \label{fig:ad_I_n}
\end{figure}

To compare the impact of the intermolecular coupling strength on the polarization-dependent harmonic yield for the adiabatic approximation with the exact results, the yield-maximizing polarization direction according to the approximation "adia-intra" is plotted in Fig.\,\ref{fig:ad_peakphis} color-coded as a function of $t_1$, similar to Fig.\,\ref{fig:polar_and_peakphis_exact}(b).\\\indent
We find that the observed behavior of the yield-maximizing laser polarization angle as a function of harmonic order and intermolecular coupling strength can be reproduced qualitatively by the adiabatic approximation: For a very weak coupling between the molecules, all harmonics maximize for $\varphi = \alphamol$ and with increasing $|t_1|$, first the maximum of the highest and then of the lower harmonic orders flip to $\varphi = \alphainter$ one after another. This means that the described flipping effect can already be explained qualitatively by just considering the adiabatic eigenstates of the system.\\\indent
However, quantitative deviations from the exact results occur. The coupling strength values at which the flipping of the yield-maximizing polarization angle happens according to the exact calculations are indicated by white dashes in Fig.\,\ref{fig:ad_peakphis}. The comparison reveals that these values are overestimated by the approximation, especially for the higher orders. We have seen that the assumption of adiabatically following electrons does not reproduce the enhanced yield around the HOMO-LUMO gap as transitions are neglected. This feature is especially important for the (high-order) intermolecular contribution, which is thus underestimated.
This effect can be seen directly in Fig.\,\ref{fig:ad_I_n}, where the harmonic intensities as a function of $t_1$ are shown for the laser polarizations perpendicular to the characteristic axes of the target using the adiabatic approximation "adia-intra".
The dashed lines, depicting the intermolecular contribution, especially those for the harmonics around and above the HOMO-LUMO gap, are clearly shifted to lower intensities compared to the exact results displayed in Fig.\,\ref{fig:t1-dep_I_n}.\\\indent
Nevertheless, the slope of the curves is reproduced correctly. Moreover, the spacing between the curves is again much smaller than for the molecular contribution (solid lines). 
The latter contribution also shows an underestimated yield for harmonic orders close to the HOMO-LUMO gap, but the overall behavior for this polarization direction, with constant intensities within a wide $t_1$-range and an ordered and approximately equally spaced distribution of the curves, is reproduced well.
This also results in the regular and ordered appearance of the crossing points between molecular and intermolecular contribution for the different harmonic orders (indicated by the diamonds), as it was observed in the exact results. However, the intersections are shifted towards stronger intermolecular couplings due to the underestimation of the intensities for $\varphi \perp \alphamol$.\\\indent
Thus, the approximation of adiabatically following electrons is able to reproduce the polarization dependence of the harmonic yield and its behavior as a function of $t_1$ with the flipping of the yield-maximizing polarization direction qualitatively correct, albeit not quantitatively.
In the condensed matter picture, this would correspond to the flipping mechanism being already encoded in the intraband contribution to the harmonic yield.

\section{Conclusion} \label{section:conclusion}
The influence of the laser polarization direction and the intermolecular coupling strength on the high-harmonic generation in systems of weakly coupled molecules was examined using a tight-binding model for a molecular dimer.
The harmonic emission shows strong low-order harmonics when the laser field is mainly polarized in the direction of the molecular axes, whereas for laser polarizations driving along the intermolecular direction, the yield is strongest around the frequency corresponding to the HOMO-LUMO gap, as transitions play a crucial role for this intermolecular contribution.
As a consequence, lower harmonic orders exhibit their maximum yield when the laser polarization angle is equal to the angle of the molecular axes, but the intensities of higher harmonic orders maximize for a polarization direction that connects the centers of the two molecules.\\\indent
The frequency at which the yield-maximizing polarization angle flips from the molecular to the intermolecular direction successively shifts to lower (higher) harmonic orders when increasing (decreasing) the intermolecular coupling strength $|t_1|$.
This is caused by a different scaling of the molecular and intermolecular HHG contribution with $t_1$.\\\indent
We find that the polarization dependence of the harmonic intensities, as well as the impact of the intermolecular coupling strength on the harmonic emission, can be reproduced qualitatively by assuming that the electrons follow the laser field adiabatically. However, the intermolecular contribution is underestimated in this approximation as transitions between the adiabatic states are neglected.\\\indent
Our results are qualitatively in line with findings in \cite{Wiechmann2025}, where high-harmonic generation in organic molecular crystals was studied. 
The fact that the main effects observed in the crystals are reproduced by our simple model system demonstrates that the model can give useful insights into HHG in systems of weakly coupled molecules in general. Due to its simplicity and flexibility, the model not only provides a framework for isolating the relevant physical mechanisms, but has also been applied to gain direct insights into the coupling-dependent electron dynamics driving the HHG process.\\\indent
In future research, the model could be advanced, for example by making it three-dimensional and using more sites per molecule. With this, it could mimic systems more realistically. To match the model to a specific real system, quantum chemistry calculations can be used to infer the tight-binding parameters for the target structure.

\begin{acknowledgments}
We acknowledge funding from the Deutsche Forschungsgemeinschaft via SFB 1477 'Light–Matter Interactions at Interfaces' (project no. 441234705).
\end{acknowledgments}


\bibliography{bib}

\end{document}